\documentclass[fp,twocolumn]{jpsj3}
\usepackage[dvipdfmx]{graphicx}
\usepackage{txfonts}
\usepackage{amsmath}
\usepackage{bm}
\usepackage{multirow}
\usepackage{url}
\usepackage{algorithmic}
\usepackage{algorithm}
\title{Systematic and Efficient Construction of Quadratic Unconstrained Binary Optimization Forms for High-order and Dense Interactions}

\author{Hyakka Nakada$^{1,2}$\thanks{hyakka\_nakada@r.recruit.co.jp} and Shu Tanaka$^{2,3,4,5}$}
\inst{$^1$Recruit Co., Ltd., Tokyo 100-6640, Japan \\
$^2$Graduate School of Science and Technology, Keio University, Kanagawa 223-8522, Japan \\
$^3$Department of Applied Physics and Physico-Informatics, Keio University, Kanagawa 223-8522, Japan \\
$^4$Keio University Sustainable Quantum Artificial Intelligence Center (KSQAIC), Keio University, Tokyo 108-8345, Japan \\
$^5$Human Biology-Microbiome-Quantum Research Center (WPI-Bio2Q), Keio University, Tokyo 108-8345, Japan}

\abst{Quantum Annealing (QA) can efficiently solve combinatorial optimization problems whose objective functions are represented by Quadratic Unconstrained Binary Optimization (QUBO) formulations. For broader applicability of QA, quadratization methods are used to transform higher-order problems into QUBOs.
However, quadratization methods for complex problems involving Machine Learning (ML) remain largely unknown. In these problems, strong nonlinearity and dense interactions prevent conventional methods from being applied.
Therefore, we model target functions by the sum of rectified linear unit bases, which not only have the ability of universal approximation, but also have an equivalent quadratic-polynomial representation. In this study, the proof of concept is verified both numerically and analytically.
In addition, by combining QA with the proposed quadratization, we design a new black-box optimization scheme, in which ML surrogate regressors are  inputted to QA after the quadratization process.}

\begin{document}
\maketitle

\section{Introduction}
\label{section:introduction}
Combinatorial optimization problems have numerous real-world applications, spanning diverse fields including logistics, materials science, and finance.
In recent years, the size of these problems has increased with the volume of data traffic, which leads to the difficulty of solving them in a realistic time. 
To address this situation, quantum computing technology has been actively developed. 
Especially, Quantum Annealing (QA)~\cite{FINNILA1994343QA, Kadowaki1998Quantum, Das2008qa, tanaka-book, tanahashi2019application, QAreview2023} is a promising metaheuristic to obtain good solutions to combinatorial optimization problems efficiently. 
Some combinatorial optimization problems can be described by either the Ising model or Quadratic Unconstrained Binary Optimization (QUBO) formulations. The Ising model and QUBO are equivalent via a linear transformation of their variables.

However, combinatorial optimization problems are generally formulated as High-Order Binary Optimization Problems (HOBO)~\cite{HOBO2022, HOBO2023} because numerous real-world phenomena exhibit strong nonlinearities. 
Thus, methods of quadratization~\cite{QD2002,QD2019} from HOBO to QUBO are widely studied to enable QA to handle such problems. 
Rosenberg's method~\cite{Rosenberg1975} is known to be one of the typical quadratization. More specifically, a product of two binary variables $x_i, x_j \ (\mathrm{for} \ i,j = 1,2,\ldots,N)$ is replaced with an auxiliary binary variable $y_{ij}$ along with a penalty term $-\lambda(x_i x_j-2x_i y_{ij}-2x_j y_{ij} + 3y_{ij})$ for a maximization problem. 
Here, $\lambda>0$ and $y_{ij} \in \{0,1\} $ denote the penalty coefficient and the auxiliary variable, respectively. $N$ is the size of the problem.
For sufficiently large $\lambda$, the penalty term enforces the binary variable assignments such that the constraint $y_{ij} = x_i x_j$ is satisfied. 
This quadratization is iterated until all of the higher-order interactions are reduced to a quadratic polynomial.
Several studies~\cite{QD2002,QD2024} have attempted to reduce the number of auxiliary variables by arranging the order of this quadratization.
\begin{figure*}[h!]
\centering
\includegraphics[width=15cm]{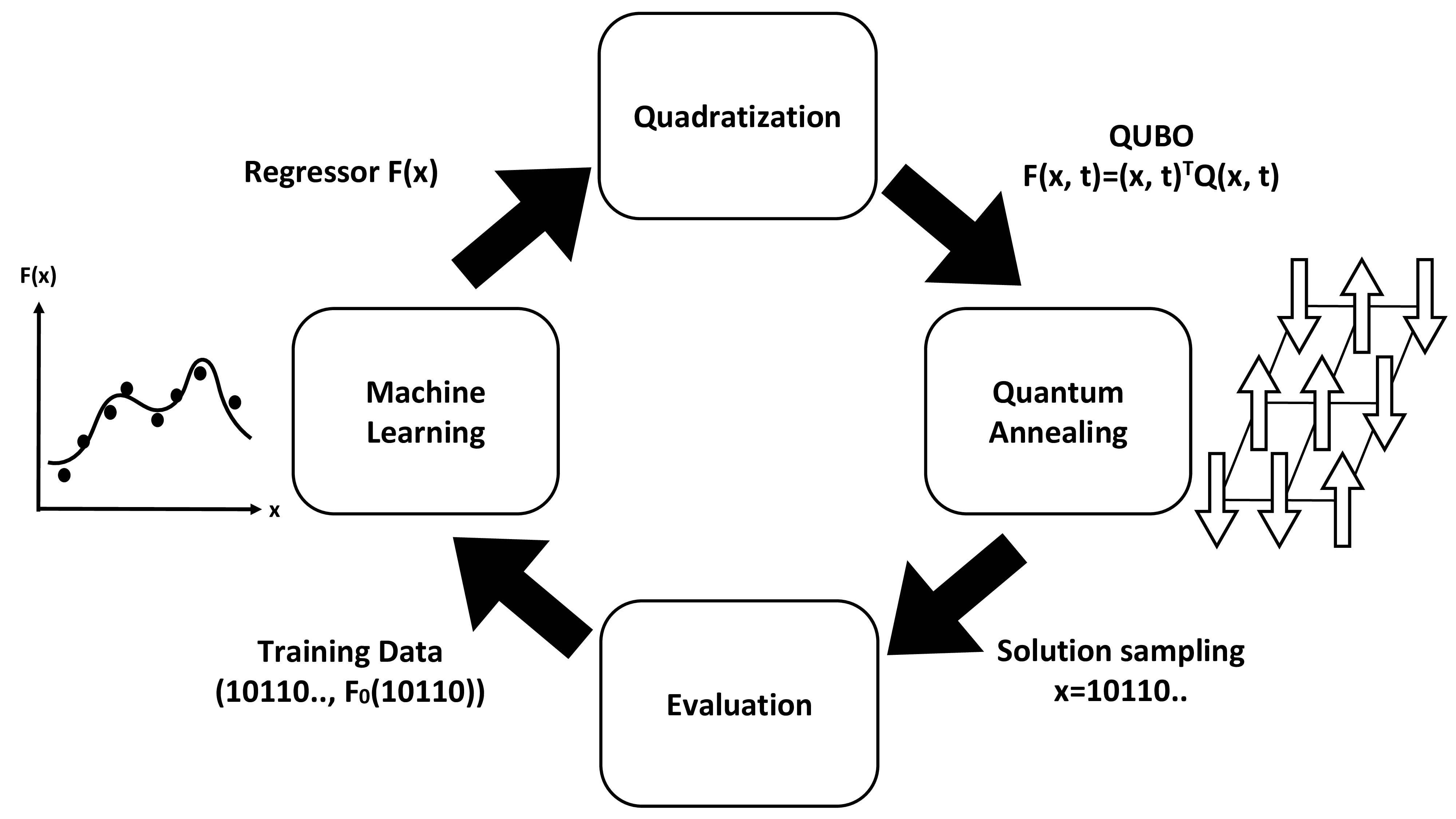}
\caption{(Color online) Basic concept of QA-based black-box optimization by using ML surrogate regressors. Whereas conventional methods adopt ad-hoc QUBO regressors that can be directly input to QA, the ML regressors are usually higher-order nonlinear functions $F(\bm{x})$, which requires quadratization process after training. Originally nonlinear regressors are quadratized to $F(\bm{x},\bm{t})$ by adding auxiliary variables $\bm{t}$. Here, $(\bm{x},\bm{t})^T$ means transpose of a row vector in which $\bm{x}$ and $\bm{t}$ are concatenated, and $Q$ is a QUBO matrix.}
\label{fig:figure1}
\end{figure*}

Recently, much attention has been paid to the handling of strong nonlinearities in the field of Machine Learning (ML), such as surrogate regressors for a value function in reinforcement learning~\cite{RL2018} and an acquisition function in Bayesian optimization~\cite{BO2012}. 
For these black-box optimizations, good candidates can be predicted by searching for maximum points along the regressors. 
Finding such points is difficult when the black-box optimization treats a vast exploration space.
QA has the potential to accelerate the optimization by efficiently searching for these candidates. 
However, because QA is specialized for handling QUBOs, the process of black-box optimization should be designed as shown in Fig.~\ref{fig:figure1}.
In other words, a quadratization step is essential to transfer trained regressors into QUBO forms because ML models are usually higher-order nonlinear.
For example, they include non-polynomial basis such as Rectified Linear Unit (ReLU) function~\cite{ReLU2011} and Radial Basis Function (RBF) kernel~\cite{KR2014}.

On the other hand, related studies on annealing-based black-box optimizations adopt surrogate regressors of a quadratic polynomial~\cite{FMA2020, FMA2022_1, FAM2022_2, FMA2023, FMA2024, KernelQA2025}.
In these methods, the regressors can be directly inputted to QA without the quadratization step.
However, as shown before, numerous real-world phenomena exhibit strong nonlinearities.
For more practical and broader applicability, ML regressors with higher-order nonlinearity are desired as well as quadratic polynomials.
Thus, as shown in Fig.~\ref{fig:figure1}, we propose a new black-box optimization method.
First, an ML regressor is trained with the available data.
Then, the regressor is transferred into a QUBO form with a quadratization method.
New candidates are selected with respect to an acquisition function based on the QUBO, by QA. 
Next, the scores of these candidates $\bm{x}$ are evaluated.
Training data $(\bm{x},F_0(\bm{x}))$ are obtained and the regressor is retrained.
Here, $F_0$ means a true but unknown objective function to describe scores.
These steps are repeated until the target score is achieved.

The important issue in realizing this optimization is quadratization.
Fortunately, in the discrete space $\bm{x}\in X=\{0,1\}^N$, many regressors can be reduced to a polynomial after expansion in a series thanks to $x^k=x$ for any natural number $k$. 
For example, $e^{x_1+x_2+x_3}=1+(e-1)(x_1+x_2+x_3)+(e-1)^2 (x_1x_2+x_2x_3+x_3x_1)+(e-1)^3 x_1x_2x_3$.
However, since the number of monomials grows exponentially with the problem size $N$, directly encoding them to HOBO is difficult in a large problem.
Similarly, encoding ML models is generally challenging.
Even if they could be encoded, the subsequent transfer to QUBO becomes intractable with existing quadratization methods involving Rosenberg’s method because the quadratization is performed sequentially on each monomial.
Therefore, in this study, the efficient quadratization step for such ML regressors is focused.

The remainder of this paper is organized as follows. Section~\ref{section:related} provides a brief summary of related studies. Section~\ref{section:proposed} proposes our quadratization approach for dense and highly nonlinear functions involving ML regressors. 
Section~\ref{section:application} describes application to several typical ML regressors.
Finally, Section~\ref{section:conclusion} concludes this paper.

\section{Related Work}
\label{section:related}
In this study, we propose a systematic and efficient QUBO formulation for functions with high-order and dense interactions involving ML regressors. 
Our target is maximizing
\begin{equation}
\label{eq:1}
F(\bm{x}) = \sum^K_{k=1} c_k f(q_k(\bm{x})).
\end{equation}
Here, we assume that $q_k:\bm{x} \to R^1$ is a scalar map and $f$ is a continuous nonlinear function.
As shown later, many ML regressors are modeled by this type of function.
As a standard technique for polynomial reduction, Piecewise-Linear Approximations (PLAs)~\cite{PLA2010} are well-established. However, PLAs for high-dimensional problems handled by ML remain largely unknown.
Although ML models are not specifically targeted, several attempts to obtain QUBO formulations for activation functions have been reported.
Previous studies~\cite{ReLUQD2019, WD2020} showed that ReLU functions
\begin{equation}
\label{eq:2}
R(q) \equiv
\begin{cases}
    \begin{aligned}
    &0\ \text{for}\ q< 0,\\
    &x\ \text{for}\ q\geq 0
    \end{aligned}
\end{cases}
\end{equation}
can be transformed to 
\begin{equation}
\label{eq:3}
R(q) = \max_{t\in [0,1]} \{tq\}.
\end{equation}
$t$ is the dual variable introduced through the Legendre transformation~\cite{LT2012}, and ReLU functions become quadratic with respect to $q,t$.

\section{Proposed Method}
\label{section:proposed}
We propose the following two quadratization methods for ML models.
The first is to linearly discretize original functions by one-hot vectorization.
The second is to expand the function using ReLU functions.
\subsection{Discretization method}
\label{section:discretization}
If $q$ is discrete, $f$ can be rigorously decomposed into $f(q)\to\sum^L_{l=1}f(d_l)s_l$ by PLAs~\cite{PLA2010}.
Here, we consider that $q$ is the $L$-ary variable and $d_l$ is the constant value at the $l$-th level. $L$ represents the total number of levels. The auxiliary variables $s_l\ \text{for} \ l=1,\ldots,L$ are one-hot vectors.
\begin{align}
\label{eq:4}
&f(q)= \notag \\
&\max_{s_l\in\{0,1\}} \left\{\sum^L_{l=1}f(d_l)s_l - \lambda \left(\sum^L_{l=1}d_l s_l-q\right)^2 - \lambda' \left(\sum^L_{l=1}s_l-1\right)^2\right\}
\end{align}
is obtained. Equation~\eqref{eq:4} is quadratic with respect to $q$ and $s_l\ \text{for} \ l=1,\ldots,L$. We refer to this as the discretization method. 
In this method, maximizing Eq.~\eqref{eq:1} is 
\begin{align}
\label{eq:5}
\max_{\bm{x}} F(\bm{x}) =& \max_{\bm{x}} \max_{\bm{s}\in\{0,1\}^{KL}} \sum^K_{k=1}\left\{\sum^L_{l=1}c_k f (d_{k,l})s_{k,l} \right. \notag \\
& \left.- \lambda_k \left(\sum^L_{l=1}d_{k,l} s_{k,l}-q_k(\bm{x})\right)^2 - \lambda'_k \left(\sum^L_{l=1}s_{k,l}-1\right)^2\right\}.
\end{align}
In this case, it is not necessary to use annealing-based solvers.
Equation~\eqref{eq:5} can be regarded as an integer linear programming problem, because the first term is linear, and the second and third terms are originated from linear constraints. In addition, several reports recently showed that such problems can be solved by tensor networks~\cite{TN2023, TN2024}.

In the discretization method, the number of auxiliary variables increases linearly with respect to $L$. In addition, adjusting the penalty coefficients $\lambda, \lambda'$ is difficult because ill-adjustments lead to invoking infeasible solutions or shifting optimal solutions.
Therefore, we propose an alternative method for QUBO formulation in the following subsections.

\subsection{ReLU-expansion method}
\label{section:relu_exp}
\begin{figure}[t!]
\begin{center}
\includegraphics[width=6.8cm]{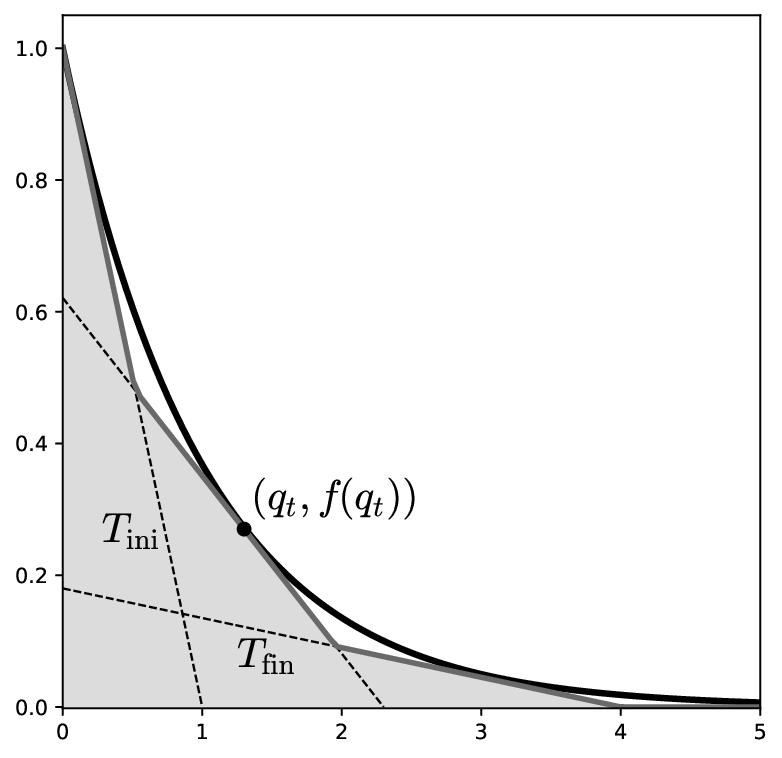}
\caption{(Color online) Black line $f(q)$ is approximated by a gray polyline. As an example, $f(q)=e^{-q}$ is depicted. The polyline is modeled by connecting the tangent lines (dashed lines).}
\label{fig:figure2}
\end{center}
\end{figure}

Our idea is that $f$ is approximated by the linear sum of a $q$-space basis that has an equivalent quadratic representation. 
When a non-polynomial basis is adopted, this approximation is justified by the universal approximation theorem~\cite{UAT1993}, which states that neural networks (NNs) with one hidden layer can approximate continuous functions on compact sets with any desired precision.
Specifically, we propose the ReLU-expansion method in which a ReLU-type basis is adopted.
Here, $f$ is approximated by a polyline as shown in Fig.~\ref{fig:figure2},
\begin{equation}
\label{eq:6}
f(q) \simeq \hat{f}(q)=\sum_{m=0}^{M-1} (a_m q +b_m) I_{\alpha_m \leq q <\alpha_{m+1}},
\end{equation}
where $I_{\alpha \leq q <\beta}$ is the indicator function that takes the value of $1$ for $\alpha \leq q <\beta$ and $0$ otherwise.
Let $M$ be the number of pieces and let $\alpha$ increase monotonically, in other words, $\alpha_0<\alpha_1<\cdots< \alpha_M$.
We assume the continuity of the right-hand side in Eq.~\eqref{eq:6} at $q=\alpha_1,\ldots,\alpha_{M-1}$ (i.e., $a_m \alpha_m+b_m =a_{m-1} \alpha_{m-1} +b_{m-1}$ for $m=2,3,\ldots,M-1$).
Equation~\eqref{eq:6} is rewritten by the sum of ReLU functions~\eqref{eq:2}, 
\begin{equation}
\label{eq:7}
\hat{f}(q)= a_0 q +b_0 + \sum_{m=1}^{M} (a_m-a_{m-1})R(q-\alpha_{m}).
\end{equation}
Here, we let $a_{M}=0$.
Thus, when the polylines along $f$ are obtained, we can derive the ReLU-expanded approximation through Eq.~\eqref{eq:7}.

A pseudocode for the ReLU-expansion method
is presented in Algorithm~\ref{alg1}.
We propose to numerically determine $a_{m}$, $b_{m}$, and $\alpha_{m}$, and there are several candidates of the procedures.
For example, these parameters can be optimized by minimizing the integration of the residual errors in Eq.~\eqref{eq:6}. 
Specifically, if $f$ is downward convex, the tangent polyline is always below $f$, as shown in Fig.~\ref{fig:figure2}. Thus, the shaded area is maximized for goodness of fitting.
For another example, one can use first-Spline interpolation~\cite{Spline1978} to derive the polylines.
Note that the values of parameters can be estimated analytically for specific types of functions as shown in Appendix~\ref{section:c2}.
\begin{figure}[!t]
\begin{algorithm}[H]
\caption{ReLU-expansion method}
\label{alg1}
\begin{algorithmic}[1]
    \renewcommand{\algorithmicrequire}{\textbf{Input:}}
    \renewcommand{\algorithmicensure}{\textbf{Output:}}
    \REQUIRE $M, D, F(\bm{x})=\sum^{K}_{k=1}c_k f(q_k(\bm{x}))$
    \ENSURE QUBO objective function
    \STATE Model approximated $f(q)$ by a polyline $\hat{f}(q)$ (Eq.~\eqref{eq:6})
    \STATE Define residual error between $f(q)$ and $\hat{f}(q)$
    \STATE Tune model parameters by minimizing the residual error
    \STATE Rewrite $\hat{f}(q)$ in ReLU form of Eq.~\eqref{eq:7} to obtain approximate objective function~\eqref{eq:8}
    \FOR{$k =1,2,\cdots,K$}
      \FOR{$m =1,2,\cdots,M$}
        \IF{$c_k(a_{m}-a_{m-1})>0$} 
         \STATE Perform Legendre transformation~\eqref{eq:3} for each ReLU function in Eq.~\eqref{eq:8} by introducing dual variables $t_{k,m}$
        \ELSE
         \STATE In Eq.~\eqref{eq:8}, 
        each Relu function is transformed by Eq.~\eqref{eq:13}, by introducing binary variables $z_{k,m,j}$ for $j=1,\ldots,D$
        \ENDIF
      \ENDFOR
    \ENDFOR
    \STATE Approximate objective function~\eqref{eq:14} is outputted
\end{algorithmic}
\end{algorithm}
\end{figure}
Thus, the target function is approximated by
\begin{align}
\label{eq:8}
F(\bm{x}) &\simeq \sum_{k=1}^K c_k\hat{f}(q_k(\bm{x}))\\ \notag
&= \sum_{k=1}^K c_k(a_0q_k+b_0)+\sum_{k=1}^K \sum_{m=1}^M c_k(a_{m}-a_{m-1})R(q_k-\alpha_{m}).
\end{align}
Maximizing Eq.~\eqref{eq:8} is 
\begin{align}
\label{eq:9}
\max_{\bm{x}} F(\bm{x})& \simeq
\max_{\bm{x}} \sum^K_{k=1}\Biggl\{ c_k a_{0} q_k(\bm{x}) \notag \\
&+ \sum^{M}_{m=1} c_k  (a_{m}-a_{m-1})\max_{t_{k,m}\in [0,1]}t_{k,m}(q_k(\bm{x})-\alpha_{m}) \Biggr\},
\end{align}
by using Eq.~\eqref{eq:3}.
Here, the constant term of Eq.~\eqref{eq:8} is omitted because it does not change the set of optimal solutions. 
\begin{figure*}[t!]
  \begin{minipage}[b]{0.67\columnwidth}
    \centering
    \includegraphics[width=5.8cm]{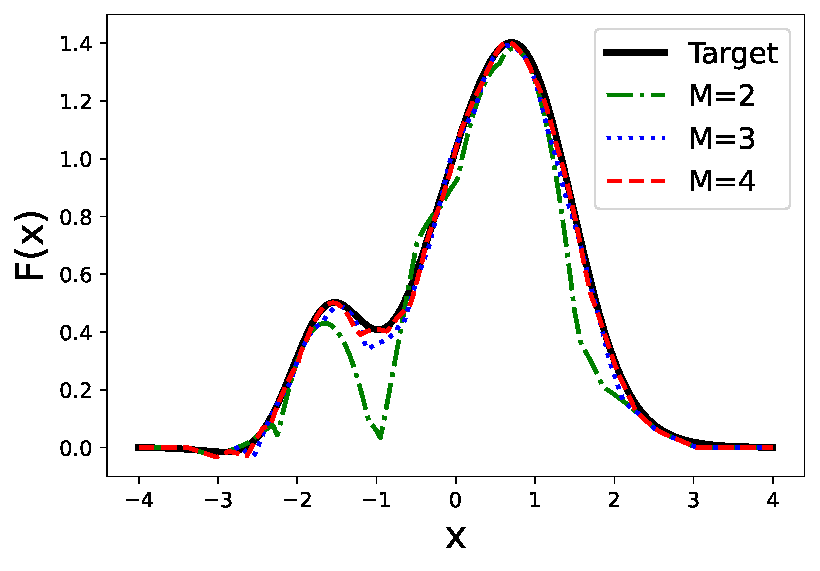}
  \end{minipage}
  \begin{minipage}[b]{0.67\columnwidth}
    \centering
    \includegraphics[width=5.8cm]{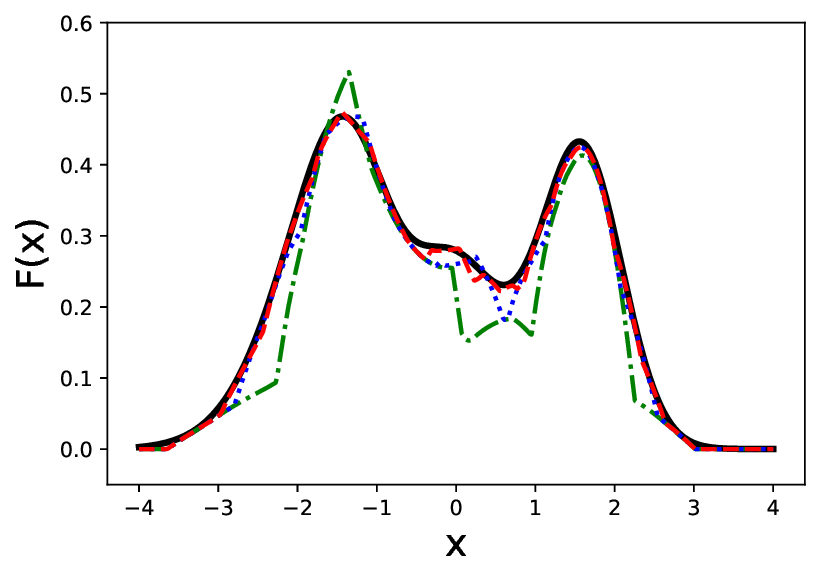}
  \end{minipage}
  \begin{minipage}[b]{0.67\columnwidth}
    \centering
    \includegraphics[width=5.9cm]{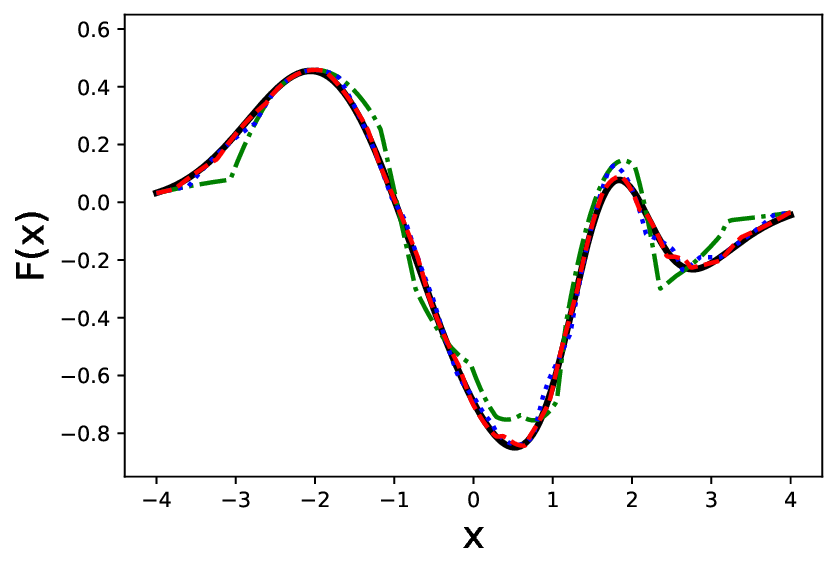}
  \end{minipage}
\caption{(Color online) Target function $F(x)$ is approximated by the ReLU-expansion method. Approximated functions are plotted as dashed lines with the total number of pieces $M=2,3$, and $4$. $F(x)=\sum_{k=1}^5 c_k \text{exp}(-(x-\mu_k)^2/2\sigma_{k}^{2})$ is generated randomly by uniform sampling $c_k\in\{-0.5,-0.4,-0.3,\ldots,1.0\}$, $\mu_k\in\{-2.0,-1.9,\ldots,2.0\}$ and $\sigma_k\in\{0.5,0.6,\ldots,1.0\}$, respectively. With only four pieces $M=4$, strongly nonlinear functions are well approximated.}
\label{fig:figure3}
\end{figure*}

\begin{table}[t]
    \caption{Optimized parameters of approximate function for $f(q)$ in Fig.~\ref{fig:figure2}.}
    \label{table:table1}
        \centering
        \begin{tabular}{ccccc}\hline \hline
        $M$& $m$ & $a_m$ & $b_m$ & $\alpha_m$ \\
        \hline
        2 &0 & -1 & 1 & 0 \\
                           &1 & -0.0498 & 0.199 & 0.8428 \\
                           &2 & - & - & 4 \\
        \hline
        3 &0  & -1 & 1 & 0  \\
                           &1  & -0.3265 & 0.6920 & 0.4574 \\
                           &2  & -0.0498 & 0.1991 & 1.7809 \\
                           &3  & - & - & 4 \\
        \hline
        4 &0 & -1 & 1 & 0 \\
                           &1 & -0.4950 & 0.8431 & 0.3108 \\
                           &2 & -0.1959 & 0.5153 & 1.0961\\
                           &3 & -0.0498 & 0.1991 & 2.1633\\
                           &4 & - & - & 4\\
        \hline
        \end{tabular}
\end{table}

In the ReLU-expansion method, for $c_k (a_{m}-a_{m-1}) > 0$, the operator $\max_{t_{k,m}}$ in the integrand can be factored in, which leads to the total maximization operator $\max_{t_{km},x}$. On the other hand, for $c_k(a_{m}-a_{m-1}) < 0$, the max-min operation $\max_{x}\min_{t_{k,m}}$ remains due to the negative sign and cannot be handled directly by QA.

To resolve such a max-min problem, an encoding technique with Wolfe duality~\cite{WD2020} is well-known, 
\begin{align}
\label{eq:10}
&\ \ \min_{t} g(t) \ \ \text{s.t.} \ h_1(t)\leq 0, h_1(t)\leq 0 \notag \\
&\Leftrightarrow  \max_{t}\max_{u,v\geq 0} \left\{g(t)+ u h_1(t) +v h_2(t)\right\} \notag \\
&\ \text{s.t.} \ \partial_t g(t) + u \partial_t h_1(t) +v \partial_t h_2(t)= 0.
\end{align}
The minimization operator can be transferred to maximization by introducing new auxiliary variables $u \geq 0, v \geq 0$. From Eq.~\eqref{eq:7}, $g(t)=c_k (a_{m}-a_{m-1})(q_k-\alpha_{m})t$, $h_1(t)=-t$ and $h_2(t)=t-1$ are obtained. The constraints can be encoded by adding the penalty term $\lambda (\partial_t g(t) + u \partial_t h_1(t)+v \partial_t h_2(t))^2$ to the objective function. 

However, this encoding requires two continuous variables $u,v$, which must be discretized into binary variables for a QUBO formulation.
Therefore, for $c_k (a_{m}-a_{m-1}) < 0$, we adopt another approach to quadratize ReLU functions.
Let $q$ be a variable $q=A(q_{\text{int}}-\alpha)$, where $q_{\text{int}}$ is an integer variable and $A> 0$ and $\alpha$ are real constants. 
The following binarization is considered. 
\begin{equation}
\label{eq:11}
q_{\text{int}}-\lfloor \alpha \rfloor = 1-2^D+\sum_{j=0}^D 2^j z_j
\end{equation}
$\lfloor x \rfloor$ denotes the floor function to output the greatest integer less than or equal to $x$.
$D$ is the bit width, which is determined to cover the domain of $q_{\text{int}}-\alpha$. In other words, $1-2^D\leq \min(q_{\text{int}}-\lfloor \alpha \rfloor)$ and $\max(q_{\text{int}}-\lfloor \alpha \rfloor)\leq 2^D$.
Thus, 
\begin{equation}
\label{eq:12}
D= \log_2 \left(\min(-\lfloor \alpha \rfloor+\max{q_\text{int}}, 1+\lfloor \alpha \rfloor -\min{q_\text{int}})\right)
\end{equation}
is obtained. 
If $z_D=0$, the right-hand side of Eq.~\eqref{eq:11} is always lower than $0$. Thus, $q/A=q_{\text{int}}-\alpha \leq q_{\text{int}}-\lfloor \alpha \rfloor \leq 0$ is obtained.
Otherwise, the right-hand side is higher than $1$, which leads to $q_{\text{int}}-\alpha > q_{\text{int}}-\lfloor \alpha \rfloor-1 \geq 0$.
This ensures that the binary variable $z_D$ works as the sign of $q$.
For an extension to the case where $q$ is not an integer variable, refer to Appendix~\ref{section:noninteger}.

Therefore, ReLU functions~\eqref{eq:2} are rewritten in the form of
\begin{equation}
\label{eq:13}
R(q) = z_Dq,
\end{equation}
under the constraint~\eqref{eq:11}.
In this study, for negative coefficients $c_k (a_{m}-a_{m-1})<0$ in Eq.~\eqref{eq:8}, $R(q_k-\alpha_m)$ is transferred by Eq.~\eqref{eq:13}. 
On the other hand, for $c_k (a_{m}-a_{m-1})>0$, Eq.~\eqref{eq:3} is utilized.

Let $\Delta_{p}$ be the index set for $(k,m)$, where $c_k(a_{m}-a_{m-1}) > 0$. We define the complement set as $\Delta_{n}$. 
After substituting Eqs.~\eqref{eq:3} and \eqref{eq:13} into Eq.~\eqref{eq:8} and rearranging the terms, the following expression is obtained.
\begin{align}
\label{eq:14}
\max_{\bm{x}} & F(\bm{x}) \simeq \max_{\bm{x}} \max_{t_{k,m}\in [0,1]} \max_{z_{j,k,m}\in \{0,1\}}
\Biggl\{ \sum^{K}_{k=1} c_k a_{0} q_k(\bm{x}) \notag \\ 
&\left. + \sum_{(k,m)\in \Delta_p} c_k(a_m-a_{m-1})t_{k,m}(q_k(\bm{x})-\alpha_{m}) \right. \notag \\ 
&\left. +\sum_{(k,m)\in \Delta_n} \Biggl\{c_k(a_m-a_{m-1})z_{k,m,D}(q_k(\bm{x})-\alpha_{m}) \right. \notag \\
&-\lambda_{k,m} \Biggl(q_k(\bm{x})-\lfloor \alpha_m \rfloor - 1+2^D-\sum_{j=0}^D 2^j z_{k,m,j}\Biggr)^2 \Biggr\}\Biggr\}
\end{align}
The last term is the penalty fucntion for Eq.~\eqref{eq:11}.
Equation~\eqref{eq:14} is the maximization of the quadratic forms with respect to $t_{k,m}$ and $z_{k,m,j}$ for $k=1,\ldots,K$, $m=1,\ldots,M$, $j=1,\ldots,D$. 
If $q_k$ is a linear map, Eq.~\eqref{eq:14} becomes quadratic also with respect to $\bm{x}$.

To achieve a QUBO formulation, $t_{k,m}$ must be discretized into binary variables. 
Fortunately, when $q\neq 0$, Eq.~\eqref{eq:3} is reduced to 
\begin{equation}
\label{eq:15}
R(q) = \max_{t\in \{0,1\}} \{tq\} \ \ \text{for} \ q\neq 0.
\end{equation}
While the previous studies~\cite{ReLUQD2019, WD2020} adopted Eq.~\eqref{eq:3}, Eq.~\eqref{eq:15} can eliminate the degree of freedom required for QUBO formulations.
Thus, $t_{k,m}$ becomes a binary variable when $q_k(\bm{x})-\alpha_{m}$ does not take the value of $0$.
In this case, $t_{k,m}$ does not require to be discretized into binary variables.

\subsection{Numerical test for ReLU-expansion method}
\label{section:test_relu_exp}
Because each of Eqs.~\eqref{eq:3} and \eqref{eq:13} is equivalent transformation itself, the exactness of the ReLU-expansion method depends mainly on the residual error in Eq.~\eqref{eq:9}. Thus, for verification, we numerically investigated the fidelity to $F(\bm{x})$.
A target function is $F(x)=\sum_{k=1}^5 c_k e^{-(x-\mu_k)^2/2\sigma_k^2}$ and the parameters $c_k$, $\mu_k$, and $\sigma_k$ are generated randomly.
Setting $q_k(x)=(x-\mu_k)^2/2\sigma_k^2$ and $f(q_k)= e^{-q_k}$, the approximate function $\hat{f}$ in Eq.~\eqref{eq:6} is estimated. 

Figure~\ref{fig:figure2} shows the approximated target functions for $M \leq 4$. 
The polyline parameters for the initial and final tangent lines $T_{\text{ini}}, T_{\text{fin}}$ were designed to pass through points $(0,1)$, and $(4,0)$, respectively. 
The parameters for the middle tangent lines were optimized by maximizing the shaded area with sequential least quadratic programming~\cite{nocedal1999numerical}, in the following procedure. 
Parameterizing the coordinates of tangent points by $q_t$, the shaded area was calculated after the middle tangent lines were analytically derived.
This area was maximized along $q_t$ to obtain the optimal tangent points, by SciPy SLSQP~\cite{2020SciPy-NMeth} with default parameters.
The optimized values are summarized in Table~\ref{table:table1}.

The approximate function was obtained by substituting the optimized parameters into Eq.~\eqref{eq:8}.
In the case of $M=2$, the approximate function was obtained by connecting only $T_{\text{ini}}$ and $T_{\text{fin}}$. 
Figure~\ref{fig:figure3} shows the target function and the approximate functions.
As the total number of pieces $M$ increases, the fidelity is improved. 

\section{Application to ML Regressors}
\label{section:application}
Using the discretization method and the ReLU-expansion method, typical ML models are encoded to QUBO. We compare their costs, such as the number of auxiliary variables and that of penalty terms as shown in Table~\ref{table:table2}.
\begin{table}[t]
    \caption{Comparison of the numbers of required auxiliary binary variables and comparison of the numbers of required penalty terms. These values are represented by $\#$A and $\#$P, respectively. $N$ denotes the problem size, in other words the dimension of $\bm{x}$. $M$ and $K$ denote the total number of pieces in the ReLU-expansion method and that of clusters of a GMM. $K^{'}=K^{'}_{p}+K^{'}_{n}$ is the number of training data in a KR, of which $K^{'}_{p}$ data have positive coefficients and $K^{'}_{n}$ data have negative ones in Eq.~\eqref{eq:18}.
    $K^{''}=K^{''}_{p}+K^{''}_{n}$ is the total number of hidden layer nodes in an NN, of which $K^{''}_{p}$ nodes have positive coefficients and $K^{''}_{n}$ nodes have negative ones in Eq.~\eqref{eq:19}.}
    \label{table:table2}
        \centering
        \begin{tabular}{ccccc}\hline \hline
        & & Discretization & ReLU exp. & Mixed \\
        \hline
        \multirow{2}{*}{GMM} &\#A & $(N+1)K$ & $MK$ & - \\
                             &\#P & $2K$ & $0$ & - \\
        \hline
        \multirow{2}{*}{KR} &\#A & $(N+1)K^{'}$ & $MK^{'}_{p}+MK^{'}_{n}D$ & $MK^{'}_{p}+(N+1)K^{'}_{n}$  \\
                            &\#P & $2K^{'}$ & $MK^{'}_{n}$ & $2K^{'}_{n}$ \\
        \hline
        \multirow{2}{*}{NN} &\#A & - & $K^{''}_{p}+K^{''}_{n}D$  & - \\
                            &\#P & - & $K^{''}_{n}$  & - \\
        \hline
        \end{tabular}
\end{table}

\subsection{Gaussian mixture model}
\label{section:gmm}
The first problem is Gaussian Mixture Models (GMMs)~\cite{GMM2009},
\begin{equation}
\label{eq:16}
F_{\text{GMM}}(x) = \sum^K_{k=1} p_k (\pi/\sigma_k^2)^{-N/2}e^{-| \bm{x}-\bm{\mu}_k|^2/2\sigma_k^2}.
\end{equation}
Here, we consider the discrete space $\bm{x}\in \{0,1\}^N$. $K$ is the total number of clusters.
Then $p_k>0$ denotes the weight in the $k$-th cluster and satisfies $\sum^K_{k=1}p_k=1$.
$\sigma_k>0$ and $\bm{\mu}_k$ denote the variance parameter and the mean vector in the $k$-th cluster, respectively. We assume $\bm{\mu}_k \in \{0,1\}^N \ \text{for} \ k=1,\ldots,K$. Equation~\eqref{eq:16} is generalized to 
\begin{equation}
\label{eq:17}
F_{\text{GMM}}(\bm{x}) = \sum^K_{k=1} c_k e^{-| \bm{x}-\bm{\mu}_k|^2/2\sigma_k^2},
\end{equation}
where $c_k \ \text{for}\ k=1,\ldots,K$ are positive coefficients. 
First, the discretization method is applied. The possible values of $q_k(\bm{x})=|\bm{x}-\bm{\mu}_k|^2/2\sigma^2_k$ are $0/2\sigma^2_k,1/2\sigma^2_k,\ldots,N/2\sigma^2_k$. Therefore, $L=N+1$ and $d_{k,j}=(j-1)/2\sigma^2_k$ in Eq.~\eqref{eq:5}. The total number of auxiliary binary variables is estimated by $L\times K=(N+1)K$ in the discretization method. That of penalty terms is $2K$.

The ReLU-expansion method can be applied in the same manner as the example in Fig.~\ref{fig:figure3}.
Because the coefficients $c_k(a_{m}-a_{m-1})$ can be assumed to be non-negative according to Appendix~\ref{section:c2}, the introduction of auxiliary variables $\bm{z}$ by Eq.~\eqref{eq:11} is not required.
Thus, Eq.~\eqref{eq:14} becomes quadratic with respect to $\bm{x},t_{k,m}$ because $q_k(\bm{x})=|\bm{x}-\bm{\mu}_k|^2/2\sigma^2_k = \sum^N_{i=1}((1-2\mu_{k,i})x_i +\mu_{k,i})/2\sigma^2_k$ is a linear map of $\bm{x}$.
Although $t_{k,m}$ is originally the continuous variable, we can assume $t_{k,m}\in \{0,1\}$ according to Eq.~\eqref{eq:15}. This is because the value of $\sigma_k^2, \alpha_{m}$ are generally decimals with large digits, which leads to $q_k(\bm{x})-\alpha_{m}=0/2\sigma^2_k-\alpha_{m},1/2\sigma^2_k-\alpha_{m},\ldots,N/2\sigma^2_k-\alpha_{m} \neq 0$. Thus, the objective function becomes the QUBO form. The total number of auxiliary binary variables is estimated by $MK$. The total number of penalty terms is $0$.

\subsection{Kernel regressor}
\label{section:kr}
The second problem is Kernel Regressors (KRs)~\cite{KR2014} and especially an RBF kernel is considered,
\begin{equation}
\label{eq:18}
F_{\text{KR}}(\bm{x}) = \sum^{K^{'}}_{k=1} c_k e^{-| \bm{x}-\bm{x}_k|^2/2\sigma^2}.
\end{equation}
Here, we consider the discrete space $\bm{x}\in \{0,1\}^N$. $K^{'}$ is the total number of training data.
$\sigma>0$ and $\bm{x}_k$ denote the variance parameter and the vector of $\bm{x}$ in the $k$-th training data, respectively. 
Because KRs have almost the same form as GMMs Eq.~\eqref{eq:17}, the total numbers of auxiliary binary variables and penalty terms in the discretization method are $(N+1)K^{'}$ and $2K^{'}$, respectively.

Next, the ReLU-expansion method is applied. 
The coefficients $c_k(a_{m}-a_{m-1})$ are not necessarily non-negative, unlike in the case of GMMs. In the following, we consider that the $K^{'}_{p}$ data have positive coefficients and the $K^{'}_{n}$ data have negative coefficients. Obviously, $K^{'}=K^{'}_{p}+K^{'}_{n}$ holds. 
For the negative coefficients, the ReLU-expansion method requires the introduction of $\bm{z}$ by Eq.~\eqref{eq:11}, leading to the increase of $D$ auxiliary variables and one constraint for each $(k,m)\in \Delta_n$.
Thus, for the negative coefficients, the numbers of auxiliary binary variables and penalty terms are $MK^{'}_{n}D$ and $MK^{'}_{n}$, respectively. The total numbers of auxiliary binary variables and penalty terms are $MK^{'}_{p}+MK^{'}_{n}D$ and $MK^{'}_{n}$. 
Because the bit width $D$ is lower than $\log_2 N$ according to Appendix~\ref{section:upperD}, the number of auxiliary binary variables is $MK^{'}_{p}+M K^{'}_n \log_2 N$ at most.

With support vector machines, KRs can be more sparse~\cite{KR2014}. In other words, the number of coefficients that take the value of $0$ in Eq.~\eqref{eq:18} increases, leading to the effective reduction of $K^{'}$.
In addition, although the RBF kernel is employed in this experiment, our method can be applied to other kernels as detailed in Appendix~\ref{section:other}. 

\subsection{Neural network}
\label{section:nn}
The third problem is NNs and especially a tri-layer network is considered,
\begin{equation}
\label{eq:19}
F_{\text{NN}}(\bm{x}) = \sum^{K^{''}}_{k=1} 
c_k R\left(\sum^N_{i=1} w_{ki} x_i +\theta_{k} \right)+c_0.
\end{equation}
Here, we consider the discrete space $\bm{x}\in \{0,1\}^N$. $K^{''}$ is the dimension of a hidden layer. $w_{ki},c_k$ are the weights and $\theta_{k},c_0$ are the biases. These values are obtained through training, to be generally decimals with large digits.
Thus, unlike the above examples, $q_k(\bm{x})=\sum^N_{i=1} w_{ki} x_i +\theta_{k}$ is not quantized.
The discretization method is difficult to apply because it requires the intractable number of auxiliary variables for $q_k$ that have exponential levels. Therefore, we consider only applying the ReLU-expansion method.

As in the case of GMMs and KRs, the input to ReLU functions is probably not $0$. Thus, the auxiliary variables $t_{k,m}$ can be reduced to binary variables.
The coefficients $c_k$ are not necessarily non-negative. In the following, we assume that the $K^{''}_{p}$ nodes have positive coefficients and the $K^{''}_{n}$ nodes have negative coefficients in Eq.~\eqref{eq:19}. 
For the negative coefficients, the ReLU-expansion method requires the introduction of $\bm{z}$, as explained in Eq.~\eqref{eq:11}.
However, the binarization~\eqref{eq:11} cannot be applied because $q_k(\bm{x})$ is generally not decomposed into $A(q_\text{int}-\alpha$).
Thus, we expand this formulation to non-integer variables by Eq.~\eqref{eq:22} in Appendix~\ref{section:noninteger}, leading to the increase of $D$ auxiliary variables and one constraint for each $k\in \Delta_n$.
Fortunately, Eq.~\eqref{eq:19} has already been decomposed into the linear sum of ReLU functions, and there are no residual errors in Eq.~\eqref{eq:7}. Thus, the total numbers of auxiliary binary variables and penalty terms are calculated by substituting $M\to1, K^{'}\to K^{''}$ into the results of KRs. In other words, the total numbers of auxiliary binary variables and penalty terms are $K^{''}_{p}+K^{''}_{n}D$ and $K^{''}_{n}$, respectively.

\subsection{Comparison between Discretization Method and ReLU-expansion Method}
\label{section:comparison}
As shown in Table~\ref{table:table2}, the numbers of required auxiliary binary variables and penalty terms are listed. Here, the mixed method means that while the ReLU-expansion method is adopted for the positive coefficients $c_k(a_{m}-a_{m-1})$, the discretization method is applied to $f(q_k(\bm{x}))$ that has the negative coefficients.

In GMMs, the ReLU-expansion method is efficient if the total number of pieces $M$ can be suppressed to less than the problem size $N$. In addition, no penalty terms exist.
On the other hand, in KRs, the numbers are highly affected by the number of negative coefficients $c_k(a_{m}-a_{m-1})$. In the optimistic case where $K^{'}_{n}=0$, the numbers of required auxiliary binary variables and penalty terms are the same as GMMs. However, in the intermediate case where $K^{'}_{n}=K^{'}/2$, the advantage of the ReLU-expansion method disappears when $M>(N+1)/\log_2 N$. Note that the ReLU-expansion method is the approximate encoding, the accuracy of which depends largely on the value of $M$. 

In NNs, the discretization method is principally difficult to be applied and only the ReLU-expansion method is considered.
The dependence of the total number of pieces $M$ vanishes because NNs are written naturally in the ReLU form.
However, although the QUBO formulation appears to be described with fewer variables than GMMs and KRs, the existence of upper limits of the bit width $D$ is not guaranteed unlike the case of them.

\section{Conclusion}
\label{section:conclusion}
In this study, we developed new quadratization methods for constructing QUBO forms for high-order and dense interactions. The first is to linearly discretize original functions by one-hot vectorization. The second is to expand the
function using ReLU functions.
We revealed that these methods can formulate QUBO from typical objective functions of ML regressors, such as GMMs, KRs, and NNs.
Then, the numbers of required auxiliary binary variables and penalty terms were estimated, which enlights the guidelines for an efficient QUBO formulation of ML models. 
As detailed in Appendix~\ref{section:other}, our method can be applied to other models.

In future work, we will try to tackle actual ML tasks. In other words, we will implement a black-box optimization with ML regressors as shown in Fig.~\ref{fig:figure1}, by using our quadratization methods. 
Because this method covers a strong nonlinearity, we would like to apply it to problems that the conventional methods adopting quadratic polynomial models cannot handle.
The proposed method has a wide variety of applications because quadratization is a fundamental preprocess prior to quantum computing technology.
For example, it can be applied to quantum approximate optimization algorithm~\cite{farhi2014qaoa, leo2020qaoa}, which was partially inspired by digitized QA. 
Thus, this study will contribute to the development of quantum gate as well as quantum annealing.

\paragraph{\footnotesize{Acknowledgments}}
\footnotesize{
This work was partially supported by the Council for Science, Technology, and Innovation (CSTI) through the Cross-ministerial Strategic Innovation Promotion Program (SIP), ``Promoting the application of advanced quantum technology platforms to social issues'' (Funding agency: QST), Japan Science and Technology Agency (JST) (Grant Number JPMJPF2221).
One of the authors S.~T. wishes to express their gratitude to the World Premier International Research Center Initiative (WPI), MEXT, Japan, for their support of the Human Biology-Microbiome-Quantum Research Center (Bio2Q).
}

\appendix
\section{Explicit ReLU-expansion for $C^2$ functions}
\label{section:c2}
Particularly, in the case of $f(q)$ is compact $C^2$, the parameters of Eq.~\eqref{eq:7} can be deduced analytically.
Let $f(q)$ be $C^2$ for $q\in [\alpha_0,\alpha_M]$. By performing integration by parts, $f$ can be rewritten in the following form,
\begin{align}
\label{eq:20}
f(q) &= f(\alpha_0)+(q-\alpha_0)f'(\alpha_0)+\int^{q}_{\alpha_0}(q-z)f''(z)dz \notag \\
&=f(\alpha_0)+(q-\alpha_0)f'(\alpha_0)+\int^{\alpha_M}_{\alpha_0}R(q-z)f''(z)dz.
\end{align}
Through piecewise quadrature, the third term is transferred into 
\begin{align}
\label{eq:21}
\int^{\alpha_M}_{\alpha_0}&R(q-z)f''(z)dz \notag \\
&=\lim_{M\to\infty} \delta \sum^{M}_{m=1} f''\left(\alpha_0+\delta m \right)R\left(q-\alpha_0-\delta m\right),
\end{align}
where $\delta=(\alpha_M-\alpha_0)/M$.
This is the infinite sum of ReLU functions.
Thus, by adopting $a_m-a_{m-1}=\delta f''(\alpha_0+\delta m)$ and $\alpha_m = \alpha_0 + \delta m$, Eq.~\eqref{eq:7} becomes a good approximation which ensures the convergence along Eq.~\eqref{eq:21}.
When $f(q)$ is concave downward (i.e., $f''(q)\geq 0$), this $(a_m-a_{m-1})$ is always non-negative, which is consistent with the results in Table~\ref{table:table1}.

\section{Generalization of Eq.~\eqref{eq:11} to non-integer variables}
\label{section:noninteger}
In Section~\ref{section:relu_exp}, we propose to utilize Eq.~\eqref{eq:11} for quadratizing ReLU functions, in the case where $q$ can be decomposed into $A(q_{\text{int}}-\alpha)$.
We consider the case where this decomposition is not applicable. The binarization is modified to
\begin{equation}
\label{eq:22}
q \simeq \max \left(\frac{|\max{q}|}{2^D},\frac{|\min{q}|}{2^D-1}\right)\left(1-2^D+\sum_{j=0}^D 2^j z_j\right).
\end{equation}
Similarly to the case where $q_{\text{int}}$ is an integer variable, the binary variable $z_D$ works as the sign of $q$.
Thus, ReLU functions are also quadratized by Eq.~\eqref{eq:13}.
In Eq.~\eqref{eq:22}, we assume the case of interest (i.e., $\min{q}<0$ and $\max{q}>0$).
This is because ReLU functions become linear trivially in the other cases, eliminating the need for quadratization.

\section{Upper bound of bit width for Eq.~\eqref{eq:11}}
\label{section:upperD}
For encoding KRs in Section~\ref{section:kr}, the upper bound of the bit width $D$ can be estimated.
In the KR model, $q_k=|\bm{x}-\bm{x}_k|^2/2\sigma^2$ where integer variable $|\bm{x}-\bm{x}_k|^2$ ranges from $0$ to $N$. Thus, Eq.~\eqref{eq:12} becomes $\log_2 \left(\min(N-\lfloor \alpha \rfloor, 1+\lfloor \alpha \rfloor\right))$.
Here, the range of interest is $0<\alpha<N$, for the same reasons as in Appendix~\ref{section:noninteger}.
Thus, $D\leq \log_2 N$ is obtained.

\section{Remark on application to other models}
\label{section:other}
Our methods can be applied to other kernels than the RBF kernel.
For example, the rational quadratic kernel~\cite{rasmussen2005kernel} is downward convex $C^2$ with respect to $|\bm{x}-\bm{x}_k|^2$, in the same manner as the RBF kernel.
Thus, the result in Section~\ref{section:kr} is still valid.
In the case of other types of kernels, though our polyline approach can be applied to, the number of required auxiliary binary variables may have different dependencies.
This is because $(a_{m}-a_{m-1})$ is no longer necessarily non-negative.

\bibliography{reference.bib}

\begin{thebibliography}{10}

\bibitem{FINNILA1994343QA}
A.~Finnila, M.~Gomez, C.~Sebenik, C.~Stenson, and J.~Doll: Chem. Phy. Lett.
  {\bfseries 219} (1994) 343.

\bibitem{Kadowaki1998Quantum}
T.~Kadowaki and H.~Nishimori: Phys. Rev. E {\bfseries 58} (1998) 5355.

\bibitem{Das2008qa}
A.~Das and B.~K. Chakrabarti: Rev. Mod. Phys. {\bfseries 80} (2008) 1061.

\bibitem{tanaka-book}
S.~Tanaka, R.~Tamura, and B.~K. Chakrabarti: {\em Quantum Spin Glasses,
  Annealing and Computation} (Cambridge University Press, 2017).

\bibitem{tanahashi2019application}
K.~Tanahashi, S.~Takayanagi, T.~Motohashi, and S.~Tanaka: Journal of the
  Physical Society of Japan {\bfseries 88} (2019) 061010.

\bibitem{QAreview2023}
B.~K. Chakrabarti, H.~Leschke, P.~Ray, T.~Shirai, and S.~Tanaka: Philosophical
  Transactions of the Royal Society A: Mathematical, Physical and Engineering
  Sciences {\bfseries 381} (2023) 20210419.

\bibitem{HOBO2022}
A.~Glos, A.~Krawiec, and Z.~Zimbor{\'a}s: npj Quantum Information {\bfseries 8}
  (2022) 39.

\bibitem{HOBO2023}
Z.~Verch{\`e}re, S.~Elloumi, and A.~Simonetto: 2023 IEEE International
  Conference on Quantum Computing and Engineering (QCE), Vol.~1, 2023, pp.
  19--25.

\bibitem{QD2002}
E.~Boros and P.~L. Hammer: Discrete Applied Mathematics {\bfseries 123} (2002)
  155.

\bibitem{QD2019}
N.~Dattani: arXiv preprint arXiv:1901.04405  (2019).

\bibitem{Rosenberg1975}
I.~G. Rosenberg: Cahiers du Centre d ’Etudes de Recherche Operationnelle,
  {\bfseries 17} (1975) 71.

\bibitem{QD2024}
L.~Schmidbauer, K.~Wintersperger, E.~Lobe, and W.~Mauerer:   (2024) 35.

\bibitem{RL2018}
R.~S. Sutton and A.~G. Barto: {\em Reinforcement learning: An introduction}
  (MIT press, 2018).

\bibitem{BO2012}
J.~Snoek, H.~Larochelle, and R.~P. Adams: Advances in neural information
  processing systems {\bfseries 25} (2012).

\bibitem{ReLU2011}
X.~Glorot, A.~Bordes, and Y.~Bengio: In G.~Gordon, D.~Dunson, and
  M.~Dud^^c3^^adk (eds), {\em Proceedings of the Fourteenth International
  Conference on Artificial Intelligence and Statistics}, Vol.~15 of {\em
  Proceedings of Machine Learning Research}, 11--13 Apr 2011, pp. 315--323.

\bibitem{KR2014}
S.~Y. Kung: {\em Kernel Methods and Machine Learning} (Cambridge University
  Press, 2014).

\bibitem{FMA2020}
K.~Kitai, J.~Guo, S.~Ju, S.~Tanaka, K.~Tsuda, J.~Shiomi, and R.~Tamura: Phys.
  Rev. Res. {\bfseries 2} (2020) 013319.

\bibitem{FMA2022_1}
S.~Izawa, K.~Kitai, S.~Tanaka, R.~Tamura, and K.~Tsuda: Phys. Rev. Res.
  {\bfseries 4} (2022) 023062.

\bibitem{FAM2022_2}
T.~Inoue, Y.~Seki, S.~Tanaka, N.~Togawa, K.~Ishizaki, and S.~Noda: Opt. Express
  {\bfseries 30} (2022) 43503.

\bibitem{FMA2023}
K.~Nawa, T.~Suzuki, K.~Masuda, S.~Tanaka, and Y.~Miura: Phys. Rev. Appl.
  {\bfseries 20} (2023) 024044.

\bibitem{FMA2024}
Y.~Seki, H.~Nakada, and S.~Tanaka: arXiv preprint arXiv:2410.12747  (2024).

\bibitem{KernelQA2025}
Y.~Minamoto and Y.~Sakamoto: arXiv preprint arXiv:2501.04225  (2025).

\bibitem{PLA2010}
R.~Misener and C.~Floudas: Journal of optimization theory and applications
  {\bfseries 145} (2010) 120.

\bibitem{ReLUQD2019}
G.~Sato, M.~Konoshima, T.~Ohwa, H.~Tamura, and J.~Ohkubo: Physical Review E
  {\bfseries 99} (2019) 042106.

\bibitem{WD2020}
T.~Yokota, M.~Konoshima, H.~Tamura, and J.~Ohkubo: Journal of the Physical
  Society of Japan {\bfseries 89} (2020) 034801.

\bibitem{LT2012}
V.~S. Denchev, N.~Ding, S.~V.~N. Vishwanathan, and H.~Neven: ICML'12, 2012, p.
  1003^^e2^^80^^931010.

\bibitem{TN2023}
J.~Lopez-Piqueres, J.~Chen, and A.~Perdomo-Ortiz: Machine Learning: Science and
  Technology {\bfseries 4} (2023) 035009.

\bibitem{TN2024}
H.~Nakada, K.~Tanahashi, and S.~Tanaka: arXiv preprint arXiv:2409.01699
  (2024).

\bibitem{UAT1993}
M.~Leshno, V.~Y. Lin, A.~Pinkus, and S.~Schocken: Neural Networks {\bfseries 6}
  (1993) 861.

\bibitem{Spline1978}
C.~De~Boor: {\em A Practical Guide to Splines} (Applied Mathematical Sciences.
  Springer New York, 1978), Applied Mathematical Sciences.

\bibitem{nocedal1999numerical}
J.~Nocedal and S.~J. Wright: {\em Numerical optimization} (Springer, 1999).

\bibitem{2020SciPy-NMeth}
P.~Virtanen, R.~Gommers, T.~E. Oliphant, et~al.: Nature Methods {\bfseries 17}
  (2020) 261.

\bibitem{GMM2009}
D.~A. Reynolds: Encyclopedia of biometrics {\bfseries 741} (2009).

\bibitem{farhi2014qaoa}
E.~Farhi, J.~Goldstone, and S.~Gutmann: arXiv preprint arXiv:1411.4028  (2014).

\bibitem{leo2020qaoa}
L.~Zhou, S.-T. Wang, S.~Choi, H.~Pichler, and M.~D. Lukin: Phys. Rev. X
  {\bfseries 10} (2020) 021067.

\bibitem{rasmussen2005kernel}
C.~E. Rasmussen and C.~K.~I. Williams: {\em Gaussian Processes for Machine
  Learning} (The MIT Press, 2005).

\end{thebibliography}

\end{document}